\documentclass[12pt]{iopart}
\usepackage{graphicx} 
\usepackage{siunitx}
\usepackage{url}
\usepackage{booktabs}
\usepackage{overpic}
\usepackage{hyperref}
\begin{document}

\title{On the timescales of controlled termination of tokamak plasmas}
\author{S Van Mulders$^1$, O Sauter$^1$}
\address{$^1$ École Polytechnique Fédérale de Lausanne (EPFL), Swiss Plasma Center (SPC), CH-1015 Lausanne, Switzerland}
\ead{simon.vanmulders@epfl.ch}

\begin{abstract}
The RAPTOR transport code has been applied to model how the time required for controlled discharge termination of Ohmic plasmas scales from present tokamaks like TCV and JET, to reactor-grade tokamaks like ITER and DEMO.
It is shown that ramping the plasma current $I_p$ down to 20\% of the flat-top value over a time interval $\Delta t_{ramp-down}=\tau_{LR}=L_i/R$, with \textit{non-normalized} internal inductance $L_i$ and resistance $R$ evaluated at stationary Ohmic conditions, results in an approximately self-similar peaking of the current density for the four considered tokamaks, indicating the adequacy of $\tau_{LR}$ as a relevant timescale for cross-machine comparisons, yielding $\tau_{LR} = \SI{0.033}{s}$ for TCV, $\SI{2.87}{s}$ for JET, $\SI{63.2}{s}$ for ITER and $\SI{166.9}{s}$ for DEMO.
Note that $\tau_{LR}$ is easy to evaluate, both in reactor design systems codes and on a tokamak real-time control system based on magnetic equilibrium reconstructions.
For the simulated ramp-downs with $\Delta t_{ramp-down}=\tau_{LR}$, the end-of-ramp-down \textit{normalized} internal inductance $\ell_{i3}$ is limited to values below 2, and a reversal of the boundary loop voltage is avoided.
An $I_p$ ramp-down faster than $\tau_{LR}=L_i/R$ requires a reversal of the boundary loop voltage and leads to the formation of a broad plasma layer carrying current in the direction opposite to the total plasma current, concomitant with $\ell_{i3}>2$, a central region with low magnetic shear and strongly peaked pressure profiles.
Significant reduction of plasma volume and elongation, as foreseen for ITER and DEMO reference ramp-down scenarios, is shown to counteract the reversal of current density and the $\ell_{i3}$ increase, while easing vertical stability control, potentially enabling faster $I_p$ ramp-down scenarios.
Further studies, both experimental and theoretical, should be performed to assess the feasibility of such fast termination scenarios, notably with respect to vertical position control, shape control and (resistive) beta limits.
Finally, a simple analytical model is proposed and applied to estimate the values of $\tau_{LR}$ based on 0D engineering parameters for different tokamaks and for varying assumptions regarding effective charge, Greenwald fraction and confinement quality.
\end{abstract}
\section{Introduction}
\label{sec:setup}
The controlled ramp-down of the plasma current $I_p$ (henceforth referred to as `ramp-down') has often been disregarded on present tokamaks, but presents a critical challenge for operation of reactor-grade tokamaks like ITER and DEMO.
To avoid routine usage of the disruption mitigation system, robust scenarios and control strategies are required to reduce $I_p$ from a nominal value of \SI{15}{MA} to at least \SI{3}{MA} for ITER \cite{deVries_2018}. \\
To achieve controlled termination of a burning tokamak plasma, various limits have to be considered simultaneously. We list here some of the main constraints during ramp-down of an ITER or DEMO deuterium-tritium high-performance plasma, as reported in \cite{deVries_2018,Poli_2018,Siccinio_2022,VanMulders_2024}: 
\begin{itemize}
    \item The transition out of a burning plasma, dominated by alpha heating, poses important control challenges. As the temperature reduces, non-linear effects on the radiated power should be accounted for, to ensure a controlled transition to L-mode. The timing of the HL transition is critical: while an early HL transition is desirable to reduce the plasma conductivity and hence counteract the peaking of the current density, the fast reduction of plasma stored energy can cause a radial excursion of the plasma, which should be controlled. Furthermore, early HL transitions are limited by the rate at which the fusion power can be reduced in burning plasmas. 
    \item As the plasma current density distribution becomes increasingly peaked during ramp-down, the plasma elongation has to be reduced, to maintain controllability of the vertical position. The elongation reduction rate is limited by ideal and resistive MHD limits and a series of machine-dependent shape control constraints, for example related to the admissible first wall heat load (e.g. avoiding the formation of a secondary X-point with strike points in sensitive regions \cite{Gribov_2016}) or related to coupling requirements of auxiliary heating sources (e.g. ensuring coupling of ion cyclotron heating or maintaining adequate tracking of rational safety factor surfaces by electron cyclotron heating and current drive to preempt or suppress neoclassical tearing modes \cite{Poli_2018}).  
    \item Sufficient auxiliary heating is required also after the HL transition, to reduce the associated beta drop, facilitating radial position control, and to avoid a radiative collapse, especially in the presence of heavy impurities like tungsten (from the first wall) and xenon (seeded in DEMO into the core plasma to reduce the heat flux to be handled in the scrape off layer), likely requiring real-time power balance control \cite{Sozzi_2021}.
    \item As $I_p$ reduces, the plasma density has to decay sufficiently fast to avoid density limit disruptions and to avoid uncontrolled detachment front movements.
\end{itemize}
According to \cite{deVries_2018} and \cite{Gribov_2016}, ITER will be able to perform a controlled ramp-down from \SI{15}{MA} to below \SI{1}{MA} in about \SI{60}{s}.
Recent works \cite{Boozer_2021, Boozer_2025} have questioned whether ITER can ramp-down as fast as the \SI{60}{s} mentioned in \cite{deVries_2018}, raising MHD stability issues linked to the peaking of the plasma current density. 
More optimistic ramp-down times (\SI{14.7}{s} and \SI{38.4}{s} for respectively ITER and DEMO) are found in \cite{Fitzpatrick_2025}, based on a simplified model for poloidal flux and electron heat diffusion in cylindrical geometry, assuming Ohmic plasmas.
In \cite{Fitzpatrick_2025}, a constant $q_{95}$ is maintained throughout most of the simulated ramp-down, assuming a limited plasma and rapidly reducing plasma elongation and volume as the plasma is shrunk against the outer wall. 
Such a strategy was experimentally shown to be an effective means to limit the $\ell_{i3}$ increase \cite{VanMulders_2024_AUG}, however its compatibility with ITER first wall heat load and shaping controllability constraints remains to be demonstrated.
Since the ITER blanket can sustain Ohmic power only for a few seconds when $I_p>\SI{7.5}{MA}$, the plasma should remain diverted for at least part of the ramp-down. 
For the diverted ramp-down simulations reported in \cite{deVries_2018}, it has been noted that the modeled ITER terminations follow a relatively high $\ell_{i3}$, low $q_{95}$ path in the $\ell_{i3}-q_{95}$ stability diagram with respect to cross-machine experimental data points, indicating the need for further studies of ideal and resistive MHD stability limits during ramp-down. 
\\
The RAPTOR transport code \cite{Felici_2018} has been used to model the ramp-down of TCV, JET and ASDEX Upgrade discharges \cite{Teplukhina_2017, VanMulders_2024_AUG}, successfully capturing the effect of the evolution plasma current, auxiliary heating and LCFS shaping on the current density and electron temperature and density dynamics. 
In \cite{VanMulders_2024}, RAPTOR has been applied to model ramp-down scenarios for DEMO, starting from burning plasma conditions at nominal $I_p=\SI{17.75}{MA}$. 
Panel (a) of Figure 9 in \cite{VanMulders_2024} shows the time evolution of the minimum of the parallel current density $j_{par}$, for different ramp-down rates $dI_p/dt=\SI{-50}{kA/s}$, $\SI{-100}{kA/s}$, $\SI{-150}{kA/s}$, $\SI{-200}{kA/s}$ that are maintained constant during the modeled ramp-down from $\SI{17.75}{MA}$ to $\SI{5.00}{MA}$ (with a simultaneous reduction of the elongation from $\kappa=1.65$ to $\kappa=1.4$), hence corresponding to time intervals ranging from \SI{255}{s} to \SI{64}{s}, and for different assumptions regarding L-mode confinement and timing of the H to L transition. For each of these ramp-down simulations, a reversal of the plasma current is observed in the outer region of the core plasma. Only the slowest ramp-down of \SI{255}{s}, with early HL transition at $t_{HL}=0.2\Delta t_{ramp-down}$ and the confinement assumption $H_{98y,2}=0.5$ during L-mode, is marginally close to avoiding negative $j_{par}$ during the modeled time interval. Note that for these simulations, auxiliary heating is maintained during L-mode to avoid a radiative collapse due to the assumed tungsten and xenon concentrations. 
\\
In the present work, RAPTOR ramp-down simulations are used to improve understanding on how the time needed for controlled plasma termination scales from a smaller tokamak like TCV to a large tokamak like JET and to the reactor-grade tokamaks ITER and DEMO. 
Like in \cite{Boozer_2025} and \cite{Fitzpatrick_2025}, this study mainly focuses on the current diffusion dynamics during ramp-down.
Other aspects, for example the feasibility of the required density decay rates to avoid density limit excursions and to avoid uncontrolled detachment (as assessed for ITER in \cite{Koechl_2018}), have not been considered here.
Furthermore, only Ohmic ramp-downs are considered. This is clearly a simplifying assumption, since auxiliary heating will likely be needed during a large segment of the ramp-down for ITER and DEMO \cite{deVries_2018, Koechl_2018, VanMulders_2024}.
To evaluate the full time needed for termination of a burning plasma, the additional time required for the burn exit phase, either during flat-top or ramp-down, should be accounted for, as in \cite{VanMulders_2024}, but has been disregarded in the present study. \\
A relevant timescale for the peaking of the plasma current density distribution, namely $\Delta t_{ramp-down}=\tau_{LR}=L_i/R$, is proposed in Section \ref{sec:timescales}, based on an energy balance argument derived from Poynting's theorem.
Section \ref{sec:model_setup} presents the modeling assumptions underlying the RAPTOR simulations shown in this paper. 
Modeling results are presented in Section \ref{sec:results}. 
For the various considered tokamaks, ramp-down simulations are performed over a time interval $\Delta t_{ramp-down}=\tau_{LR}$, as well as for a significantly shorter time $\Delta t_{ramp-down}=0.6\tau_{LR}$.
An analytical model to estimate $\tau_{LR}$ based on volume-averaged electron temperature $\langle T_e\rangle_{vol}$, minor radius $a$, $\ell_{i3}$, $q_{95}$ and shaping parameters, taking into account the neoclassical corrections to the plasma conductivity from \cite{Sauter_1998, Sauter_2002}, is proposed in Section \ref{sec:scalinglaw}, together with a simple power balance model to estimate $\langle T_e\rangle_{vol}$ for stationary Ohmic conditions.
The model is then applied to calculate $\tau_{LR}$ for Ohmic L-mode scenarios with varying effective charge, Greenwald fraction and confinement quality, for TCV, JET, ITER, DEMO and SPARC.
Finally, further experimental and theoretical studies are proposed to assess vertical position control, shape control and (resistive) beta limits for termination scenarios with $\Delta t_{ramp-down}<\tau_{LR}$.
\section{Relevant timescales}
\label{sec:timescales}
When the current of an electrical system is changed on a timescale that is fast compared to the penetration of magnetic flux in the conductor, the inductance of the system can be decomposed in an external inductance $L_{e}$, depending on geometric factors only, and internal inductance $L_{i}$, which depends on the internal current density distribution \cite{Romero_2010}.
The total energy of the magnetic field due to $I_p$ inside and outside of the conductor can then be written as respectively $L_{i}I_p^2/2$ and $L_{e}I_p^2/2$ \cite{Romero_2010}.
For tokamaks, the normalized internal inductance $\ell_{i3}=2L_i/(\mu_0R_0)$ has been defined. 
An increased peaking of $j_{par}$ corresponds to an increased value of $\ell_{i3}$ (see eq. (\ref{eq:Iphi_int}) in \ref{sec:Li_deriv}).
\\
The evolution of the poloidal flux in a tokamak plasma is governed by a diffusion equation that originates from Faraday's law, Ampere's law and Ohm's law in toroidal geometry.
For a radially non-uniform toroidal plasma loop voltage $U_{pl}$, this equation drives the radial redistribution of the current density.
The present paper investigates the limitations that the poloidal flux diffusion equation implies for the minimum controllable ramp-down time.
While the central solenoid allows to rapidly reduce the toroidal loop voltage at the last closed flux surface $U_{pl,b}$, the inward propagation of this voltage takes place over the characteristic timescale linked to poloidal flux diffusion, the resistive time $\tau_R=\mu_0 (a/2)^2\langle \sigma \rangle$, with plasma minor radius $a$ and volume-averaged plasma conductivity $\langle \sigma \rangle$. 
A rapid $U_{pl,b}$ reduction hence causes a transient state with strongly peaked Ohmic current density and increased values of $\ell_{i3}$.
\\
In \cite{Ejima_1982}, Poynting's theorem is applied to describe the energy balance at the last closed flux surface (LCFS) enclosing an axisymmetric toroidal plasma:
\begin{equation}
U_{pl,b}I_p = \frac{d(L_{i}I_p^2/2)}{dt} + RI^2,
\label{eq:Poynting}
\end{equation}
where we have defined the plasma resistance $R$ as $P_{oh}/I_p^2$, with the Ohmic power $P_{oh}=\int j_{tor} E_{tor}dV$.
The interpretation of this energy balance is rather straightforward: the electromagnetic energy entering the LCFS, $U_{pl,b}I_p$, is the sum of Ohmic dissipation, $RI^2$, and the rate of change of the energy of the magnetic energy generated by $I_p$ inside the LCFS, ${d(L_{i}I_p^2/2)}/{dt}$. \\
In the absence of energy influx into the LCFS, i.e. with $U_{pl,b}=0$, eq. (\ref{eq:Poynting}) describes the dissipation of the internal magnetic energy linked to $I_p$ through Ohmic heating. With $U_{pl,b}=0$, we can extract an equation relating $dL_{i}/dt$ and $dI_{p}/dt$ from eq. (\ref{eq:Poynting}),
\begin{equation}
\frac{1}{2L_i}\frac{dL_i}{dt} = -\frac{R}{L_i}-\frac{1}{I_p}\frac{dI_p}{dt}. 
\label{eq:RL}
\end{equation}
Therefore, $L_i$ increases (${dL_i}/{dt}>0$) if ${d\log I_p}/{dt}<-R/L_i$, or, equivalently, when the $I_p$ decay time $I_p/|{dI_p}/{dt}|$ is fast with respect to the characteristic time $\tau_{LR}=L_i/R$ (with $dI_p/dt<0$).
Furthermore, for constant $L_i$ and $R$ during ramp-down, eq. (\ref{eq:RL}) yields the solution $I_p(t)=I_{p,FT}\exp(-t/\tau_{LR})$, with $I_{p,FT}$ the flat-top plasma current.  
In practice, both $L_i$ and $R$ evolve due to current diffusion and electron heat diffusion in the plasma, as studied with the RAPTOR code in Section \ref{sec:results}.
We however maintain $\tau_{LR}=L_i/R$, evaluated for nominal flat-top conditions, as a useful time constant to compare against the numerical modeling results.
\\
Finally eq. (\ref{eq:Poynting}) also indicates that a faster reduction of the internal magnetic energy, and hence $I_p$, can be achieved with $U_{pl,b}<0$, using the poloidal field coils to extract energy from the plasma.
However, the resulting negative loop voltage causes the reversal of $j_{par}$ near the edge, as will be analysed further based on the modeling results in Section \ref{sec:results}.
\section{Modeling set-up}
\label{sec:model_setup}
The simulations in this paper are performed with the RAPTOR transport code \cite{Felici_2018}, evolving the time evolution of the profiles of electron temperature $T_e(\rho,t)$ and poloidal flux $\psi(\rho,t)$ by solving the non-linear, coupled partial differential equations for electron heat and current density transport (with as radial coordinate the normalized toroidal flux label $\rho = \sqrt{\Phi/\Phi_b}$ where $\Phi$ is the toroidal magnetic flux enclosed by a flux surface, and $\Phi_b$ the total toroidal flux enclosed by the LCFS). 
\begin{table}[htbp]
\centering
\caption{Comparison of stationary flat-top conditions modeled in RAPTOR for TCV, JET, ITER and DEMO.}
\label{tab:flattop}
\begin{tabular}{lcccc}
\toprule
\textbf{Variable} & \textbf{TCV} & \textbf{JET} & \textbf{ITER} & \textbf{DEMO} \\
\midrule
$I_{p}$ [MA]    & 0.300      & 2.15 & 15.00   & 17.75 \\
$a$ [m]         & 0.24       & 0.91 & 2.00       & 2.92 \\
$R_0$ [m]       & 0.88       & 2.88 & 6.20       & 8.95 \\
$B_0$ [T]       & 1.41       & 2.80 & 5.30       & 5.86 \\
$\epsilon$      & 0.27       & 0.32 & 0.32       & 0.33 \\
$\kappa$        & 1.52       & 1.63 & 1.80       & 1.72 \\
$\delta$        & 0.22       & 0.26 & 0.41       & 0.37 \\
$q_{95}$        & 2.95        & 3.81  & 2.91        & 3.46 \\
$V$ [m$^3$]     & 1.40       & 72.2 & 818     & 2410 \\
$P_{oh}$ [MW]   & 1.39       & 3.02 & 12.0       & 10.6 \\
$\tau_{E,98y,2}$ [s] & 0.029      & 1.04 & 15.8     & 31.0 \\
$\tau_E$ [s]    & 0.010      & 0.42 & 6.3      & 12.7 \\
$f_{Gw}$        & 1.00       & 1.00 & 1.00       & 1.00 \\
$\langle n_{e} \rangle_{vol}$ [$10^{19}$m$^{-3}$]      & 16.5       & 8.24 & 12.13       & 6.40 \\
$T_{e0}$ [keV]  & 0.25       & 0.98 & 2.63       & 3.50 \\
$\langle T_{e}\rangle_{vol}$ [keV]  & 0.13       & 0.41 & 1.47       & 1.68 \\
$\langle \chi_{e} \rangle_{vol}$ [m$^2$/s]  & 4.55      & 2.17 & 0.49       & 0.53 \\
$q_{95}$        & 2.95 & 3.98 & 2.96 & 3.60 \\
$\ell_{i3}$     & 0.93       & 1.04 & 0.87       & 0.98 \\
$\beta_{pol}$   & 0.38       & 0.19 & 0.13       & 0.10 \\
$\beta_N$       & 0.93       & 0.60 & 0.39       & 0.27 \\
$\tau_R$ [s]    & 0.024      & 1.80 & 43.0     & 112.6 \\
$\tau_{LR}$ [s] & 0.033      & 2.87 & 63.2    & 166.9 \\
$\tau_{R}/\tau_{LR}$ & 0.73       & 0.62 & 0.68    & 0.67 \\
$j_{par,0}$ [MA/m$^2$]   & 2.85       & 1.41 & 1.55       & 0.97 \\
$U_{pl,b}$ [V]   & 4.69       & 1.40 & 0.58       & 0.59 \\
$U_{pl,b}/(2\pi R_0)$ [V/m]   & 0.85       & 0.078 & 0.020       & 0.011 \\
\bottomrule
\end{tabular}
\end{table}
\\
Sawtooth instabilities are simulated with the models described in \cite{Porcelli_1996, Sauter_1998}, implemented in RAPTOR in \cite{Piron_2015}, triggering a sawtooth crash when the magnetic shear at $q=1$ exceeds the critical value $s_{q=1, crit}=0.2$ (this value, for Ohmic plasmas, is in accordance with previous studies \cite{Sauter_1998, Fevrier_2018, VanMulders_2026}), relaxing the $q$ profile with the Kadomtsev’s complete magnetic reconnection model \cite{Kadomtsev_1975} and flattening density and pressure are flattened within the mixing radius (as defined in \cite{Porcelli_1996}).
\\
Each ramp-down simulation is initiated from the stationary, i.e. fully relaxed, solution of $T_e$ and $\psi$, considering only Ohmic heating. Various quantities of the flat-top solution for the considered tokamaks are compared in Table \ref{tab:flattop}.
\subsection{Poloidal flux diffusion}
\label{sec:polflux}
\begin{table}[htbp]
\centering
\caption{Comparison of ramp-down times $\Delta t_{ramp-down}$, $dI_p/dt$ and $dI_{p,N}/dt_N$ for the simulations for TCV, JET, ITER and DEMO reported in this paper.}
\label{tab:rd_times}
\begin{tabular}{lccccc}
\toprule
 & \textbf{Variable} & \textbf{TCV} & \textbf{JET} & \textbf{ITER} & \textbf{DEMO} \\
\midrule
$\Delta t_{ramp-down} = \tau_{LR}$ & $\Delta t_{ramp-down}$ [s] & 0.033 & 2.87 & 63.2 & 166.9 \\
 & $\left|{dI_p}/{dt}\right|$ [MA/s] & 7.273 & 0.599 & 0.190 & 0.085 \\
\midrule
$\Delta t_{ramp-down} = 0.6\tau_{LR}$ & $\Delta t_{ramp-down}$ [s] & 0.020 & 1.72 & 37.9 & 100.1 \\
 & $\left|{dI_p}/{dt}\right|$ [MA/s] & 12.121 & 0.999 & 0.317 & 0.142 \\
\end{tabular}
\end{table}
The poloidal flux diffusion equation is solved without any auxiliary current density sources.
Neoclassical conductivity and bootstrap current are computed with the Sauter model \cite{Sauter_1999, Sauter_2002}.
The time trace of $I_p$ is prescribed as a Neumann boundary condition.
For each of the simulated ramp-downs, $I_p$ is reduced linearly from a nominal flat-top value $I_{p,FT}$ to an end-of-ramp-down value equal to 20\% of $I_{p,FT}$, i.e. $I_{p,end}=0.2I_{p,FT}$ (DEMO: $I_{p,FT}=\SI{17.75}{MA}$; $I_{p,end}=\SI{3.55}{MA}$; ITER: $I_{p,FT}=\SI{15}{MA}$; $I_{p,end}=\SI{3}{MA}$; JET: $I_{p,FT}=\SI{2.15}{MA}$; $I_{p,end}=\SI{0.43}{MA}$; TCV: $I_{p,FT}=\SI{300}{kA}$; $I_{p,end}=\SI{60}{kA}$). Since $I_{p,end}=\SI{3}{MA}$ is considered the current below which benign unmitigated disruptions are expected \cite{deVries_2018}, we define the time required to reach this current for ITER as the ramp-down time interval. For TCV, JET and DEMO, we assess the time required to reach the same $I_{p,end}/I_{p,FT}=0.2$.
For each tokamak, ramp-down simulations are performed over time intervals equal to $\Delta t_{ramp-down} = \tau_{LR}$ and $\Delta t_{ramp-down} = 0.6\tau_{LR}$, the timescale introduced in Section \ref{sec:timescales} and evaluated for stationary flat-top conditions ($\tau_{LR} = \SI{0.033}{s}$ for TCV, $\SI{2.87}{s}$ for JET, $\SI{63.2}{s}$ for ITER and $\SI{166.9}{s}$ for DEMO).
As shown in Table \ref{tab:flattop}, $0.6\tau_{LR}$ is roughly comparable to the resistive time $\tau_R$, following the definition introduced in Section \ref{sec:timescales} ($\tau_R/\tau_{LR}$ is equal to $0.73$, $0.62$, $0.68$ and $0.67$ for respectively TCV, JET, ITER and DEMO). 
The ramp-down times for the various tokamaks for both $\Delta t_{ramp-down} = \tau_{LR}$ and $\Delta t_{ramp-down} = 0.6\tau_{LR}$ are summarized in Table \ref{tab:rd_times}, as well as the corresponding $dI_p/dt$ rates.
\subsection{Electron temperature diffusion}
The electron temperature diffusion equation is solved without any heat sources or sinks due to auxiliary heating or radiation, since we consider only Ohmic ramp-downs.
At the LCFS, $T_e$ is prescribed as a Dirichlet boundary condition (\SI{50}{eV} for TCV and \SI{100}{eV} for JET, ITER and DEMO). \\
The gradient-based transport model \cite{Teplukhina_2017} has been used. Transport coefficients are calculated such that a core scale length $\lambda_{T_e}=-{\partial \log T_e}/{\partial \rho}$ and an edge average gradient $\mu_{T_e}=-{\partial T_e}/{\partial \rho}$ result for respectively core $\rho \in [0\ \rho_{ped}]$ and edge $\rho \in [\rho_{ped}\ 1]$ regions under stationary conditions.
We have assumed $\lambda_{T_e}=3$ (using the value obtained for ASDEX Upgrade and JET L-modes in \cite{Teplukhina_2017}) and $\rho_{ped}=0.8$.
The edge gradient parameter $\mu_{T_e}$ is feedback controlled in the simulation to maintain a confinement factor $H_{98y,2} \sim 0.4$, with $H_{98y,2}=\tau_E/\tau_{E,98y,2}$, with $\tau_{E,98y,2}$ the energy confinement time evaluated based on the IPB98(y,2) scaling law \cite{ITER_TC_1999} (with the Ohmic power as heating power). A confinement factor $H_{98y,2} =0.4$ is chosen, assuming L-mode confinement quality, since only Ohmic heating is considered. \\
The electron density is prescribed as a quadratic profile and rescaled during ramp-down to maintain a constant Greenwald fraction equal to one.
The present simulations hence do not assess the limitation of the density decay rate for realistic particle confinement time and pumping capacity.  
The ion temperature $T_i$ is set equal to $T_e$.
Deuterium is considered as main ion species and carbon is added as a proxy for the intrinsic impurities, with a concentration such that the ion species densities satisfy quasi-neutrality and a plasma effective charge $Z_{eff}=1.5$. \\
It is important to note that there is an impact of parameters like effective charge $Z_{eff}$, Greenwald density fraction and confinement factor $H_{98y,2}$ on the plasma resistivity and hence $\tau_{LR}$ and $\tau_{R}$. Parameter scans can be performed to assess the sensitivity of the current density evolution on these modeling assumptions.
Such a sensitivity study can also be performed based on a simplified analytical model, as illustrated in Section \ref{sec:scalinglaw}.
\subsection{Equilibrium geometry evolution}
\begin{figure}
    \centering
    \includegraphics[width=0.45\linewidth]{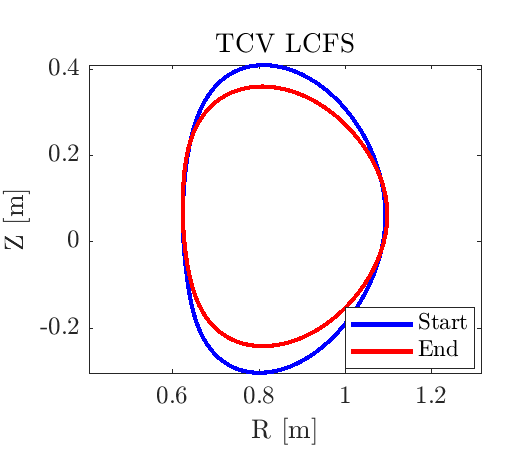}
    \includegraphics[width=0.45\linewidth]{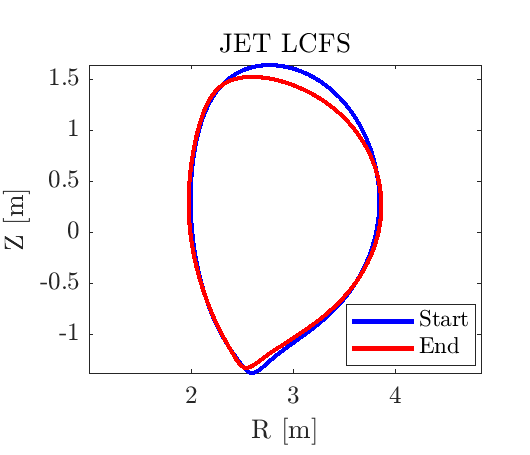}
    \includegraphics[width=0.45\linewidth]{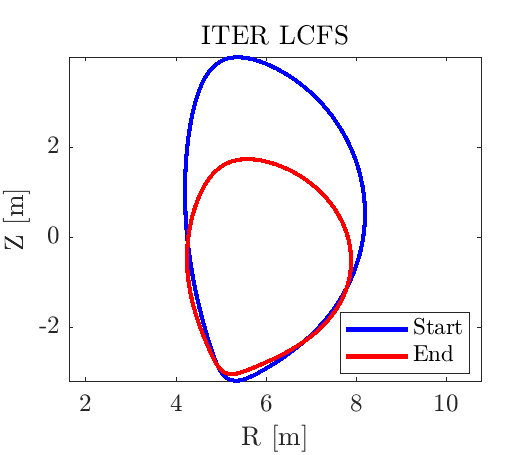}    
    \includegraphics[width=0.45\linewidth]{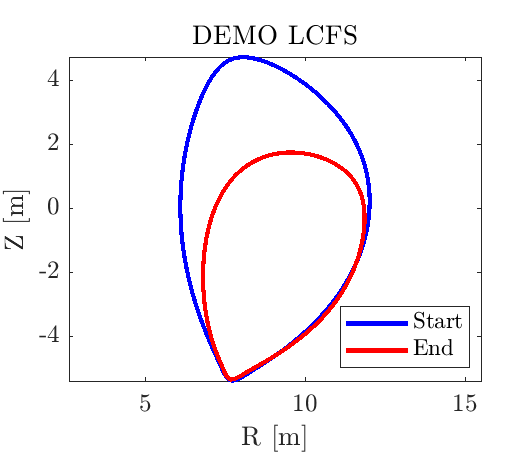}   
    \caption{LCFS evolution TCV, JET, ITER, DEMO}
    \label{fig:LCFSevol}
\end{figure}

A time-dependent equilibrium geometry can be provided to RAPTOR \cite{Teplukhina_2017}, prescribing the corresponding geometric coefficients at a set of time points and performing a linear interpolation for intermediate times. 
In the present simulations, the geometry is prescribed at the beginning and at the end of the simulated ramp-down phase.
Two cases are considered for each ramp-down: (1.) a case where the LCFS is maintained constant; (2.) a case where elongation and volume are reduced.
The equilibrium geometry is calculated with the CHEASE fixed-boundary equilibrium solver \cite{Lutjens_1996}, prescribing the LCFS shapes shown in Figure \ref{fig:LCFSevol} and providing the parallel current density $j_{par}$ and pressure $p$ obtained in RAPTOR, such that the equilibrium geometries are consistent with the internal profiles. 
For each of the simulations in this paper, the prescribed LCFS shape evolution has been either achieved in experiment (for TCV and JET) or obtained with free-boundary equilibrium simulations (for ITER and DEMO), giving some confidence that these shapes are controllable with the respective tokamak control systems.
Note however that proper assessment would require free-boundary equilibrium simulations with the same ramp-down times that have been used in this work and with the corresponding internal profile evolution obtained in RAPTOR. Such assessment is left for future studies.
\begin{table}[htbp]
\centering
\caption{Comparison of volume and elongation TCV, JET, ITER and DEMO.}
\label{tab:equil_evol}
\begin{tabular}{lcccc}
\toprule
\textbf{Variable} & \textbf{TCV} & \textbf{JET} & \textbf{ITER} & \textbf{DEMO} \\
\midrule
$V_{flat-top}$ [m$^3$] & 1.40 & 74.7 & 819 & 2411 \\
$V_{end}$ [m$^3$] & 1.20 & 71.0 & 489 & 1374 \\
$V_{end}/V_{flat-top}$ & 0.85 & 0.95 & 0.60 & 0.57 \\
\midrule
$\kappa_{flat-top}$ & 1.52 & 1.63 & 1.80 & 1.72 \\
$\kappa_{end}$ & 1.27 & 1.51 & 1.31 & 1.42 \\
$\kappa_{end}/\kappa_{flat-top}$ & 0.84 & 0.93 & 0.73 & 0.82 \\
\bottomrule
\end{tabular}
\end{table}
\\
The equilibria for ITER (panel (c) of Figure \ref{fig:LCFSevol}) are based on the ramp-down phase of a full-discharge DINA-JINTRAC simulation (shot number 134173; run number 106 in the IMAS database) of a fusion gain $Q=10$ baseline scenario, which has been discussed in \cite{Koechl_2018} and \cite{VanMulders_2025}. 
The equilibria for DEMO (panel (d) of Figure \ref{fig:LCFSevol}) are based on CREATE-NL free boundary equilibrium calculations, which have been reported in \cite{CREATE_NL} and which were used for the ramp-down optimization studies in \cite{VanMulders_2024}. 
Both the ITER and DEMO equilibria maintain a diverted lower single null configuration throughout the modeled ramp-down. Elongation and volume are significantly reduced, as quantified in Table \ref{tab:equil_evol}, while the LCFS shape close to the X-point remains mostly unchanged, minimizing changes to the magnetic geometry nearby the divertor strike points and therefore to the pumping efficiencies. \\
For TCV and JET, the shape evolution (respectively shown in panel (a) and (b) of Figure \ref{fig:LCFSevol}) is based on equilibrium reconstructions of the discharges TCV\#64965 (an Ohmic discharge, discussed in \cite{Marin_2025,VanMulders_2026}) and JET\#96432 (a hybrid scenario discharge with ion cyclotron heating and neutral beam injection, discussed in \cite{Hobirk_2023}). 
Note that for the JET discharge, the on-axis toroidal magnetic field $B_0$ reduces significantly during ramp-down.
Since such a reduction of $B_0$ is not feasible for tokamak reactors, the present simulations consider constant $B_0$. However, to illustrate the significant impact of such a reduction of $B_0$ on the evolution of $\ell_{i3}$, an additional simulation is added in Section \ref{sec:results} where this effect is modeled. Finally, it is important to note that the elongation and volume reduction for the TCV and JET scenarios are significantly smaller compared to the ITER and DEMO scenarios, as shown in Table \ref{tab:equil_evol}.
\section{Ramp-down simulations and interpretation}
\label{sec:results}
\begin{figure}
    \centering
    \includegraphics[width=\linewidth]{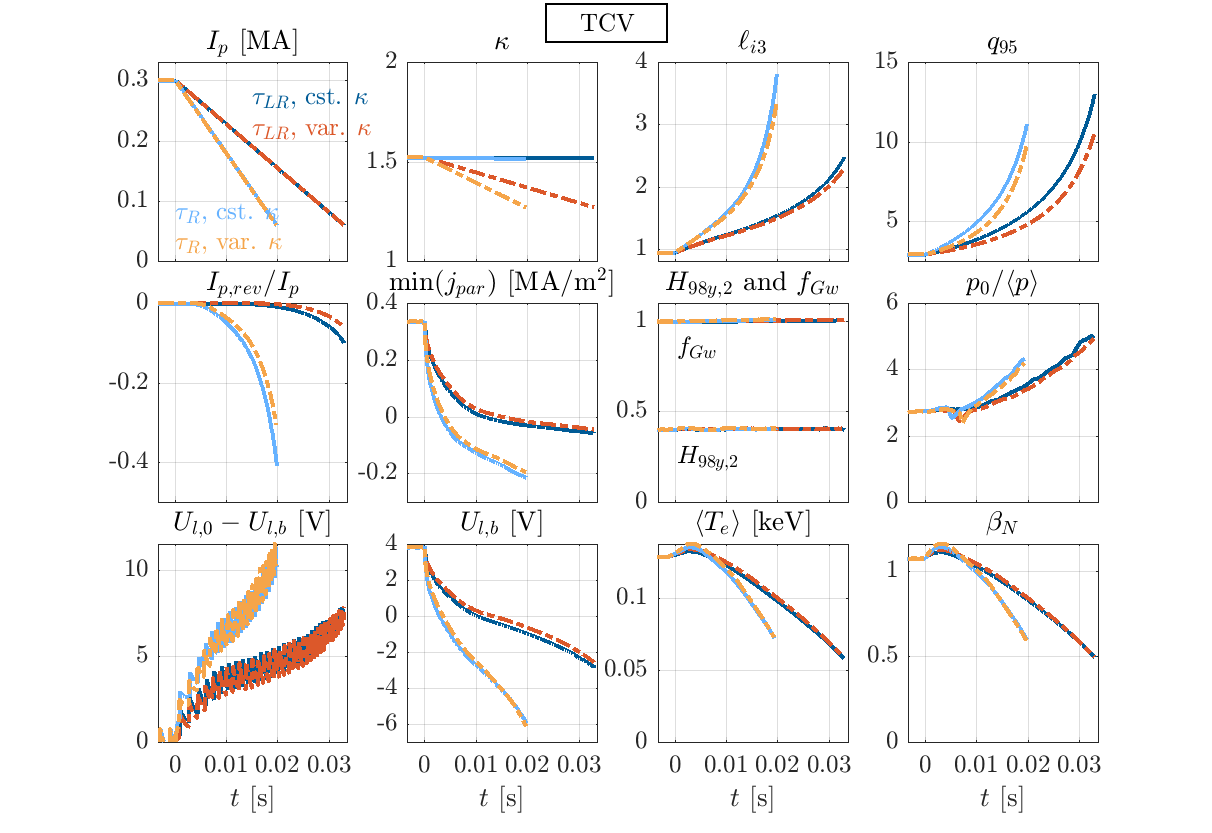}
    \includegraphics[width=\linewidth]{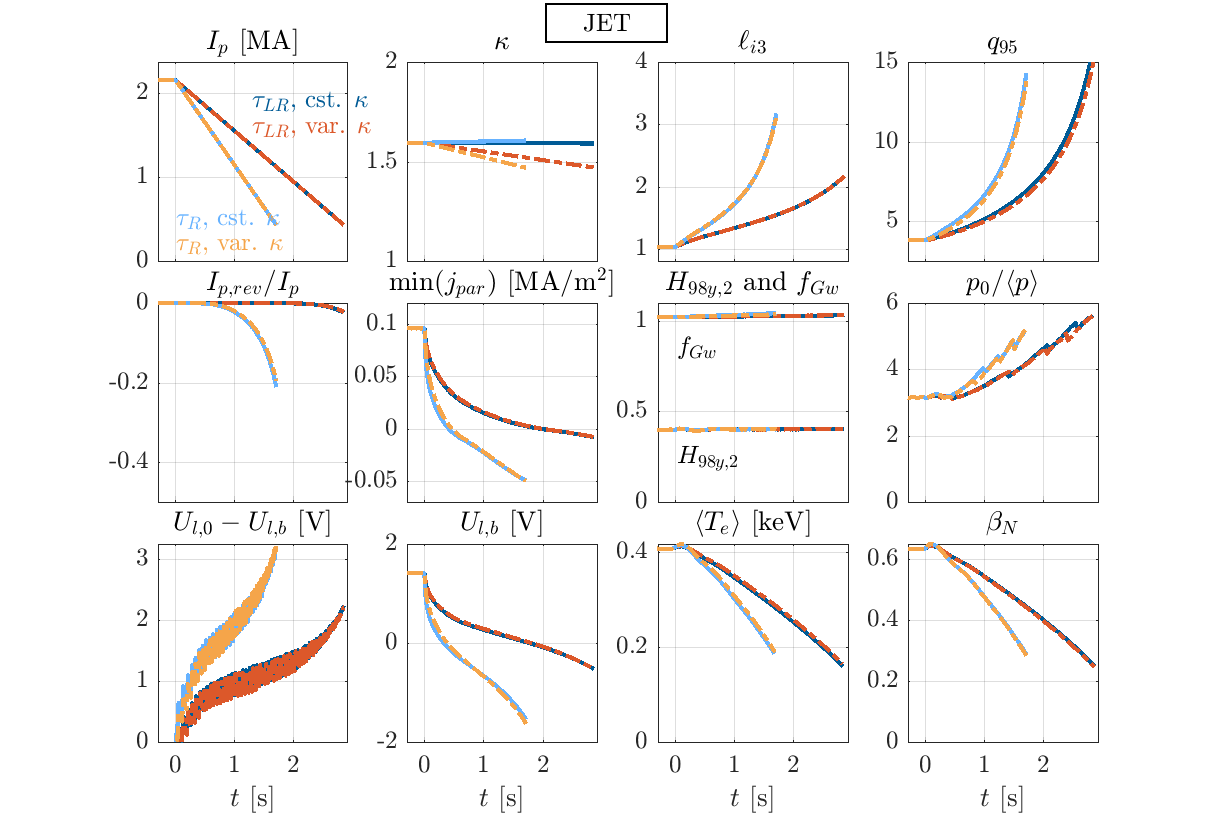}
    \caption{TCV (top) and JET (bottom) RAPTOR ramp-down simulations over $\Delta t_{ramp-down}=\tau_{LR}$ and $\Delta t_{ramp-down}=0.6\tau_{LR}$, for constant and varying $\kappa$. (a) $I_p$ [MA]; (b) $\kappa$; (c) $\ell_{i3}$; (d) $q_{95}$; (e) $I_{p,rev}/I_p$, with integrated reverse current $I_{p,rev}$; (f) $\min(j_{par})$ [MA/m$^2$]; (g) H-factor $H_{98y,2}$, Greenwald fraction $f_{Gw}$; (h) pressure peaking factor $p_0/\langle p\rangle$; (i) on-axis to boundary loop voltage difference $U_{l,0}-U_{l,b}$ [V]; (j) boundary loop voltage $U_{l,b}$; volume-averaged $\langle T_e \rangle$; (k) $\beta_N$.}
    \label{fig:TCV_JET_tauLRR}
\end{figure}
\begin{figure}
    \centering
    \includegraphics[width=\linewidth]{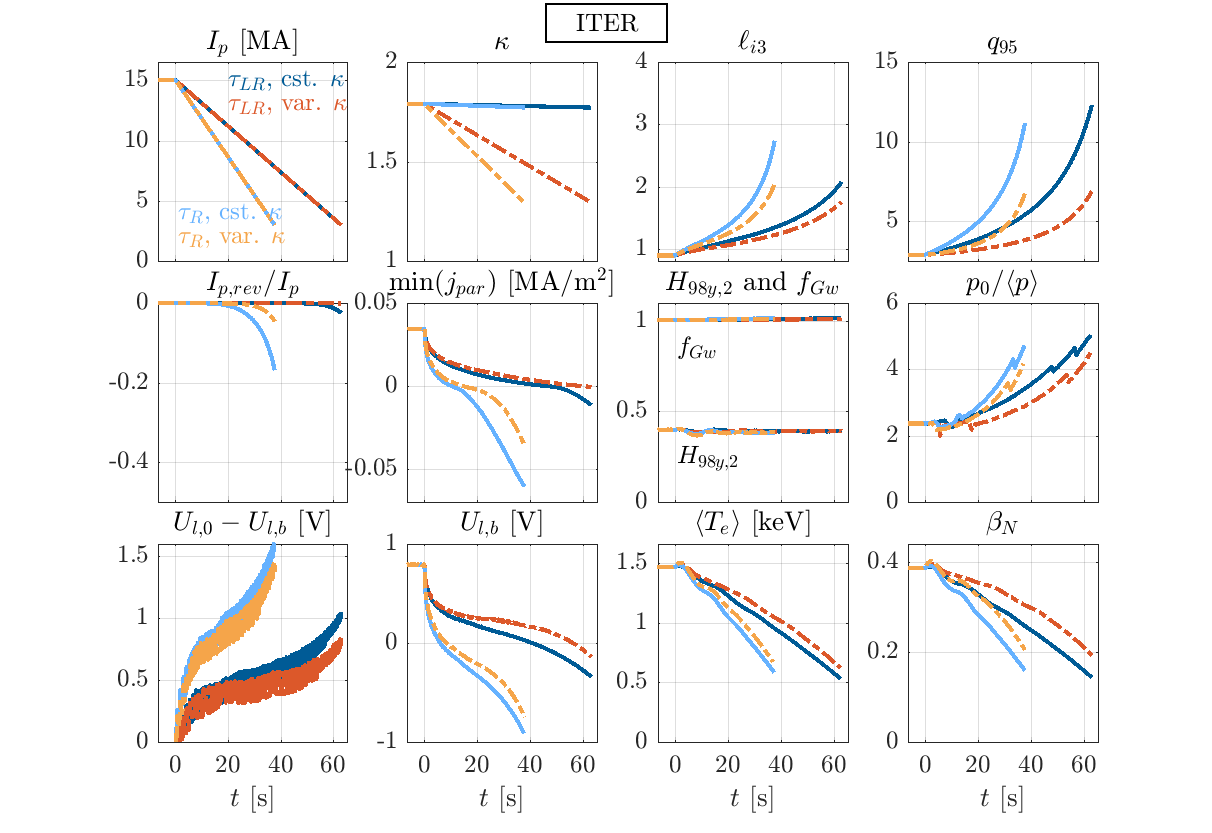}
    \includegraphics[width=\linewidth]{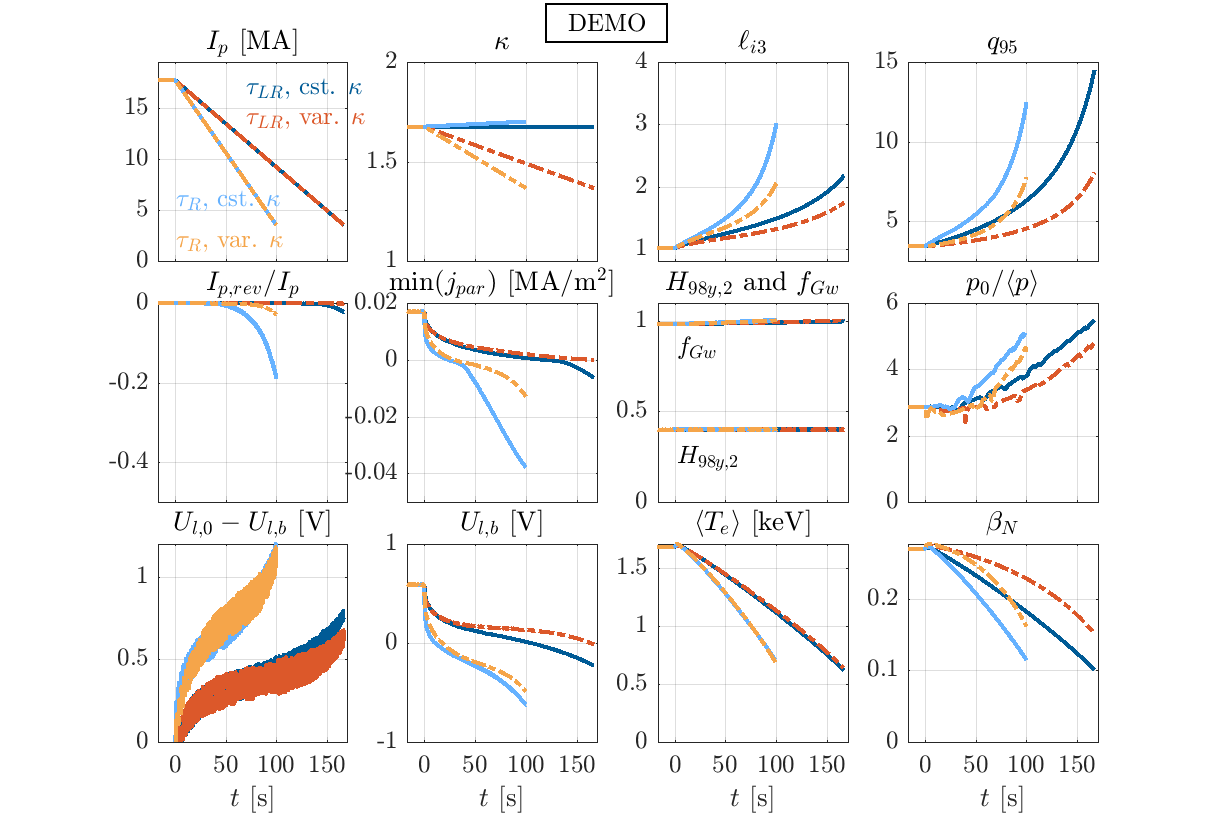}
    \caption{ITER (top) and DEMO (bottom) RAPTOR ramp-down simulations over $\Delta t_{ramp-down}=\tau_{LR}$ and $\Delta t_{ramp-down}=0.6\tau_{LR}$, for constant and varying $\kappa$. (a) $I_p$ [MA]; (b) $\kappa$; (c) $\ell_{i3}$; (d) $q_{95}$; (e) $I_{p,rev}/I_p$, with integrated reverse current $I_{p,rev}$; (f) $\min(j_{par})$ [MA/m$^2$]; (g) H-factor $H_{98y,2}$, Greenwald fraction $f_{Gw}$; (h) pressure peaking factor $p_0/\langle p\rangle$; (i) on-axis to boundary loop voltage difference $U_{l,0}-U_{l,b}$ [V]; (j) boundary loop voltage $U_{l,b}$; volume-averaged $\langle T_e \rangle$; (k) $\beta_N$.}
    \label{fig:ITER_DEMO_tauLRR}
\end{figure}
\begin{figure}
    \centering
    \includegraphics[width=0.45\linewidth]{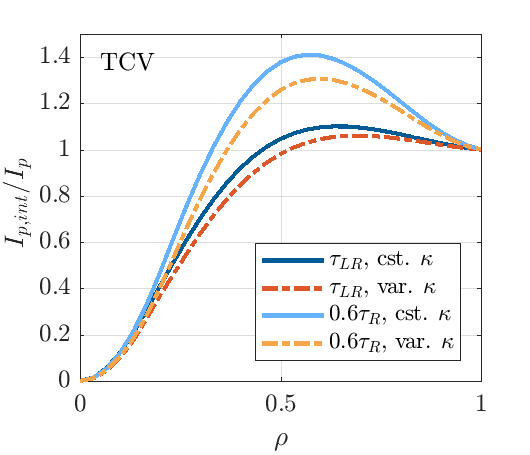}
    \includegraphics[width=0.45\linewidth]{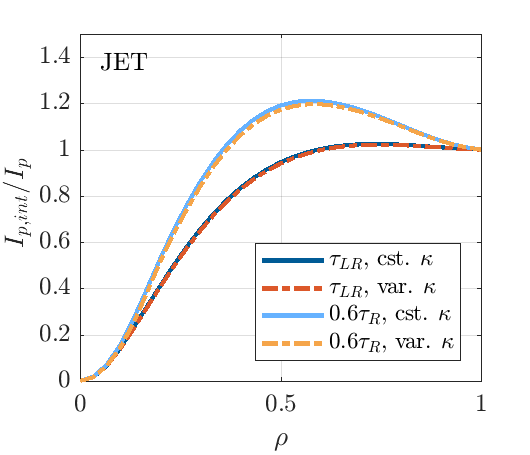}
    \includegraphics[width=0.45\linewidth]{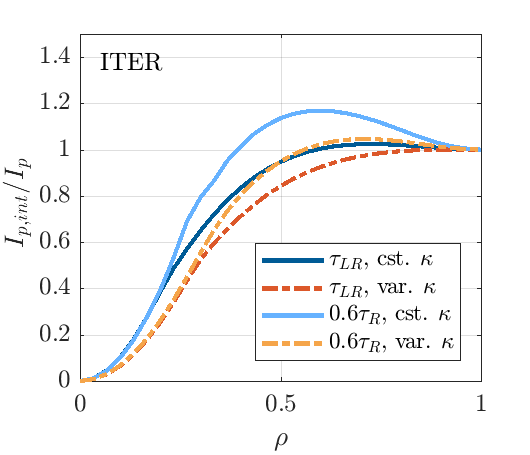}
    \includegraphics[width=0.45\linewidth]{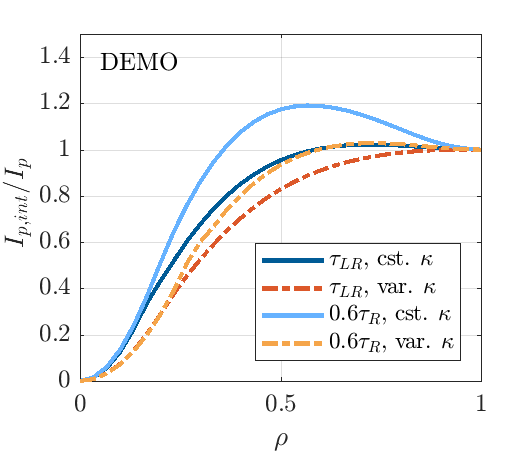}
   \caption{Normalized enclosed plasma current $I_{p,int}/I_p$ [MA] at the end of the RAPTOR simulation for TCV (top left), JET (top right), ITER (bottom left) and DEMO (bottom right), for ramp-down times $\Delta t_{ramp-down}=\tau_{LR}$ and $\Delta t_{ramp-down}=0.6\tau_{LR}$, for constant and varying $\kappa$.}
  \label{fig:Ipl_end}
\end{figure}
For all considered tokamaks, the slower termination scenarios with ramp-down time $\Delta t_{ramp-down}=\tau_{LR}=L_i/R$ lead to an increase of the internal inductance from a flat-top value around $\ell_{i3}\sim1$ to a final value of about $\ell_{i3}\sim2$, as shown in the panels (c) of Figure \ref{fig:TCV_JET_tauLRR} for TCV and JET and in Figure \ref{fig:ITER_DEMO_tauLRR} for ITER and DEMO.
These slower ramp-downs avoid the formation of significant reverse edge current, as illustrated in Figures \ref{fig:TCV_JET_tauLRR} and Figure \ref{fig:ITER_DEMO_tauLRR} by the time traces of
\begin{itemize}
    \item $I_{p,rev}/I_p$ (panel (e)): the surface integral of the toroidal current density directed opposite to the total plasma current, normalized to $I_p$;
    \item $\min (j_{par})$ (panel (f)): the time evolution of the minimum value of the parallel current density profile;
    \item $U_{l,b}$ (panel (j)): the boundary loop voltage.
\end{itemize}
When these values become negative, the plasma current density is locally reversed in the outer plasma. 
To achieve the faster $\Delta t_{ramp-down}=0.6\tau_{LR}$ ramp-down, a significant outer plasma layer with negative current density is formed. 
For the constant elongation cases, the negative current, flowing in the direction opposite to $I_p$, has a magnitude of about $20\%$ (JET, ITER, DEMO) to $40\%$ (TCV) of the total $I_p$.
This is further illustrated by the profiles of the normalized enclosed plasma current profile $I_{p,int}/I_p$, with $I_{p,int}=\int j_{\phi}dS_\phi$ as a function of $\rho$ at the end of the RAPTOR simulation, in Figure \ref{fig:Ipl_end}. 
A non-monotonic $I_{p,int}$ profile that overshoots the total plasma current $I_{p,int}(\rho=1)=I_p$ corresponds to a plasma with a reverse $j_{par}$ edge region (the overshoot in $I_{p,int}/I_p$ contributes to an increased $L_i$, see eq. (\ref{eq:Iphi_int}) in \ref{sec:Li_deriv}). 
For TCV and JET, a clear overshoot is visible for the fast ramp-downs with $\Delta t_{ramp-down}=0.6\tau_{LR}$, while $I_{p,int}$ is essentially monotonic for the slower ramp-downs with $\Delta t_{ramp-down}=\tau_{LR}$.
While a similar behavior is found for the ITER and DEMO ramp-downs with constant elongation, the fast ramp-down simulations with varying elongation successfully avoid a significant $I_{p,int}$ overshoot.
For the ITER and DEMO ramp-downs, reducing elongation and volume has a significant limiting effect on the increase of both $q_{95}$ and $\ell_{i3}$.
\begin{figure}
    \centering
    \includegraphics[width=\linewidth]{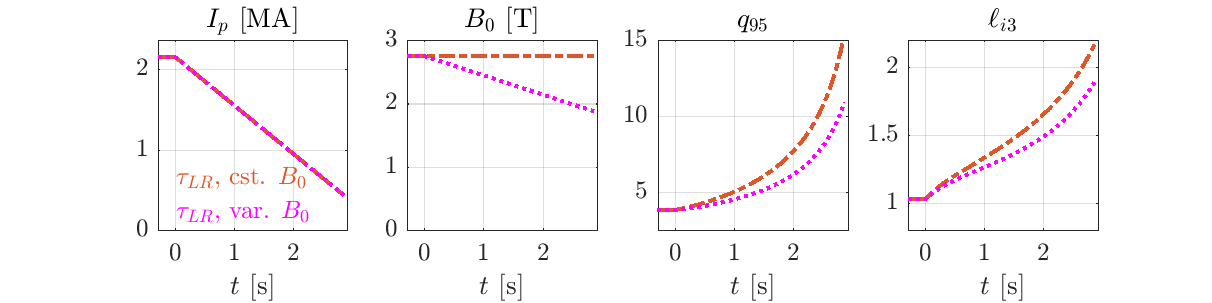}
    \caption{JET RAPTOR ramp-down simulations over $\Delta t_{ramp-down}=\tau_{LR}$ with varying $\kappa$, for constant and reducing on-axis toroidal magnetic field $B_0$. (a) $I_p$ [MA]; (b) $B_0$ [T]; (c) $q_{95}$; (d) $\ell_{i3}$.}
    \label{fig:JET_lowBt}
\end{figure}
The elongation and volume reduction have a lesser effect on $q_{95}$ and $\ell_{i3}$ for the TCV and JET cases, since the volume reduction is less significant, as quantified in Table \ref{tab:equil_evol}. 
The increase of $q_{95}$ can also be limited by reducing the toroidal magnetic field.
To illustrate this effect, we have performed an additional simulation for the JET $\Delta t_{ramp-down}=\tau_{LR}$ ramp-down, reducing the on-axis magnetic field $B_0$ from \SI{2.8}{T} to \SI{1.9}{T}.
As shown in Figure \ref{fig:JET_lowBt}, the $B_0$ reduction leads to a significant reduction of the $\ell_{i3}$ increase with respect to the reference case with constant $B_0$.
Since a reduction of $B_0$ is not foreseen for the ramp-down scenarios in ITER and DEMO, this effect has to be accounted for when comparing to termination scenarios on present-day tokamaks.
\begin{figure}
    \centering
    \includegraphics[width=0.45\linewidth]{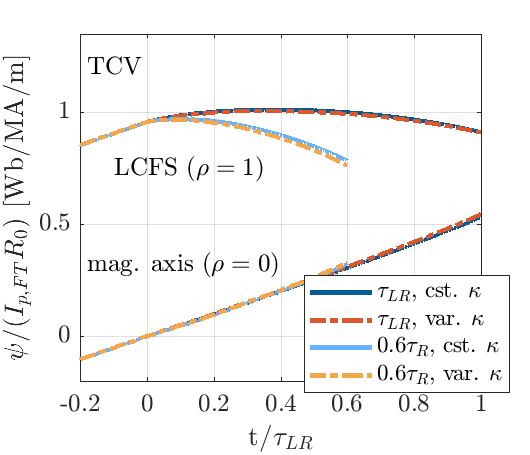}
    \includegraphics[width=0.45\linewidth]{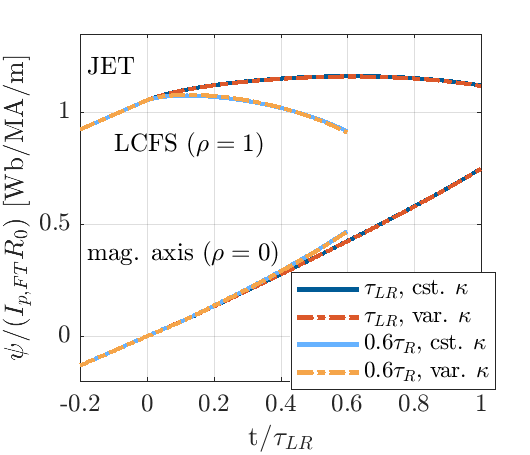}
    \includegraphics[width=0.45\linewidth]{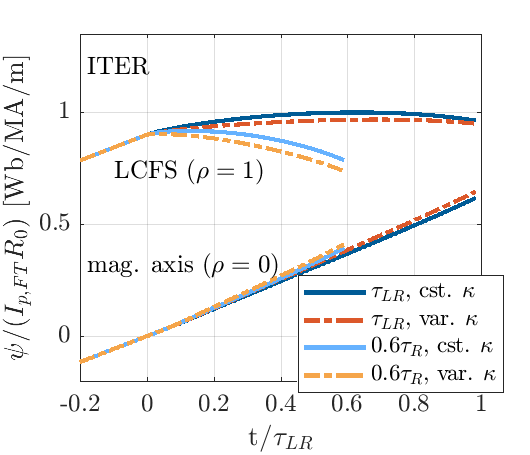}
    \includegraphics[width=0.45\linewidth]{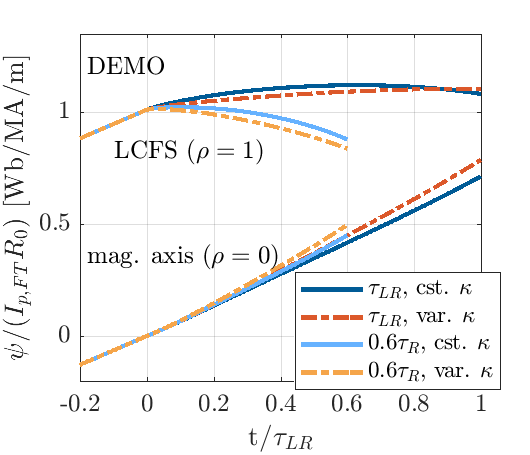}
   \caption{Poloidal flux $\psi$ normalized by flat-top $I_p$ and $R_0$ [Wb/MA/m] at $\rho=1$ (top traces) and $\rho=0$ (bottom traces) for the RAPTOR simulation for TCV (top left), JET (top right), ITER (bottom left) and DEMO (bottom right), for ramp-down times $\Delta t_{ramp-down}=\tau_{LR}$ and $\Delta t_{ramp-down}=0.6\tau_{LR}$, for constant and varying $\kappa$. Time on the abscissa is normalized to $\tau_{LR}$ for the various tokamaks. The equivalent non-normalized time is shown in Figures \ref{fig:TCV_JET_tauLRR} and \ref{fig:ITER_DEMO_tauLRR}. The $\psi$ traces are shifted such that $\psi(\rho=0)=0$ at the start of ramp-down.}
  \label{fig:diffpsi}
\end{figure}
\\
The transient nature of the current diffusion dynamics, both for the faster $\Delta t_{ramp-down}=0.6\tau_{LR}$ and the slower $\Delta t_{ramp-down}=\tau_{LR}$ ramp-down, is clearly illustrated by the time evolution of the poloidal flux at the magnetic axis, $\psi(\rho=0)$, and at the LCFS, $\psi(\rho=0)$, as shown in Figure \ref{fig:diffpsi}.
To highlight the self-similarity of the poloidal flux dynamics of these simulations for the various tokamaks, the time coordinate is normalized by $\tau_{LR}$, while the poloidal flux is normalized by $I_{p,FT}R_0$, with the flat-top plasma current $I_{p,FT}$ in MA (this normalization is justified by the derivation in \ref{sec:Li_deriv}, where it is shown that $L_{i}I_p \sim \psi_{LCFS}-\psi_{axis}$, with $L_i=\mu_0\ell_{i3}R_0/2$).
While $\psi(\rho=1)/dt$ reduces, corresponding to a lowering edge loop voltage $U_{l,b}$, the central poloidal flux continues increasing approximately linearly in time.
For each of the simulated ramp-downs, there is insufficient time for the edge loop voltage $U_{l,b}$, driven by the central solenoid, to propagate to the magnetic axis, so that $\psi(\rho=0)/dt$ and $U_{l,0}$ remain essentially unaffected.
This is also illustrated by the growing loop voltage difference $U_{l,0}-U_{l,b}$ shown in panels (i) of Figure \ref{fig:TCV_JET_tauLRR} and Figure \ref{fig:ITER_DEMO_tauLRR}.
Since a radially flat loop voltage corresponds to a stationary, relaxed current density distribution, the increase of $U_{l,0}-U_{l,b}$ is a measure for the proximity to a stationary state. 
In such fast ramp-down scenarios, peaking of the loop voltage $U_{l,0}-U_{l,b}$ is the dominant driver of the Ohmic current density peaking, while peaking of the neoclassical conductivity, hence $T_e$ profile effects, play a secondary role.
\begin{figure}
    \centering
    \includegraphics[width=1\linewidth]{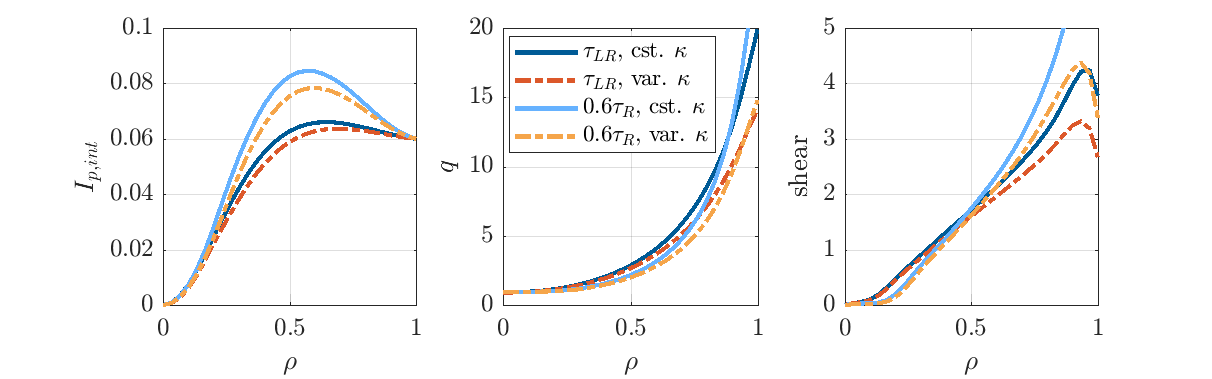}
    \caption{Enclosed plasma current $I_{p,int}$ [MA], safety factor $q$ and magnetic shear $s$ at the end of the RAPTOR simulation for TCV, for ramp-down times $\Delta t_{ramp-down}=\tau_{LR}$ and $\Delta t_{ramp-down}=0.6\tau_{LR}$, for constant and varying $\kappa$. The simulation time step right before the triggering of the last sawtooth crash is chosen, corresponding to the diffused current density profile.}
    \label{fig:TCV_profiles}
\end{figure}
\begin{figure}
    \centering
    \includegraphics[width=0.45\linewidth]{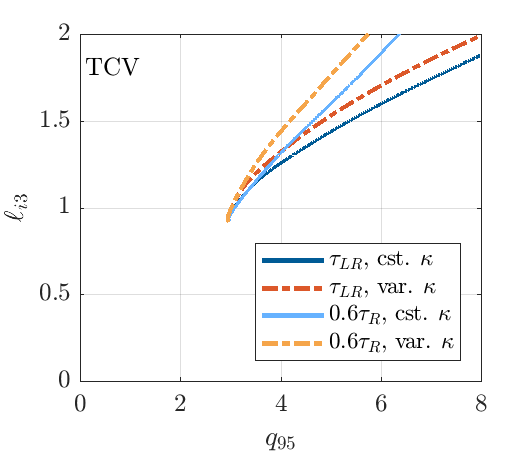}
    \includegraphics[width=0.45\linewidth]{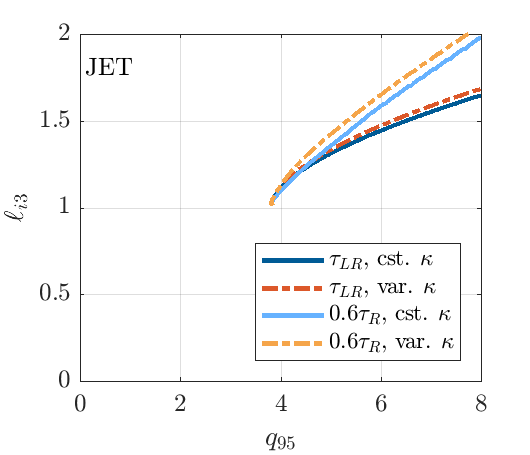}
    \includegraphics[width=0.45\linewidth]{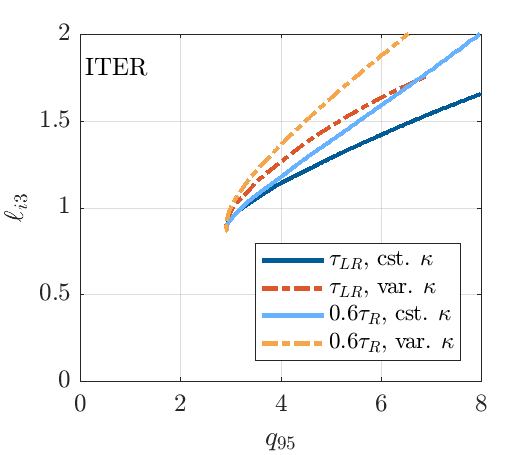}
    \includegraphics[width=0.45\linewidth]{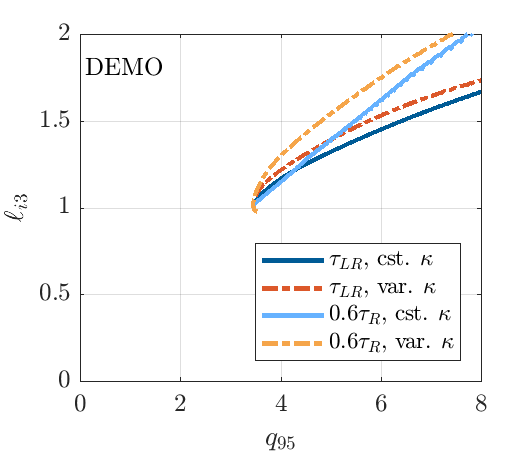}
   \caption{$(q_{95},\ell_{i3})$ evolution for TCV (top left), JET (top right), ITER (bottom left) and DEMO (bottom right) for RAPTOR ramp-down simulations over $\Delta t_{ramp-down}=\tau_{LR}$ and $\Delta t_{ramp-down}=0.6\tau_{LR}$, for constant and varying $\kappa$. The x-axis and y-axis show the same segments of $q_{95}$ and $\ell_{i3}$ as in Figure 5 of \cite{deVries_2018}, to ease comparison of the ramp-downs modeled here with the cross-machine experimental database reported there.}
   \label{fig:qli_traces}
\end{figure}
\\
The significant current density peaking of the fast TCV and JET ramp-downs and the fast, constant-elongation ITER and DEMO ramp-downs is also clear from the increase of internal inductance in excess of 2, to values around $\ell_{i3}\sim3$, which is well above the values typically observed during controlled ramp-downs (see Figure 4 and 5 in \cite{deVries_2018} and Figure 13 in \cite{Gerasimov_2020}). 
The non-monotonic $I_{p,int}$ profile at the end of these fast ramp-downs corresponds to an increased magnetic shear for $\rho>0.5$ and a lowered magnetic shear in the core, for $\rho<0.5$, as illustrated for the TCV simulations in Figure \ref{fig:TCV_profiles}. All simulated ramp-downs exhibit a significant peaking of the pressure profile, as shown by the pressure peaking factor $p_0/\langle p\rangle$, with on-axis pressure $p_0$ and volume-averaged pressure $\langle p \rangle$, in Figures \ref{fig:TCV_JET_tauLRR} and \ref{fig:ITER_DEMO_tauLRR}, with the end-of-ramp-down $p_0/\langle p\rangle$ between 4 and 5 for all cases. The peaking of the pressure profile is mainly due to the inward movement of the $q=1$ surface as $q_{95}$ rises, hence reducing the inner region affected by sawtooth flattening.
Simultaneous low magnetic shear in the plasma core and high pressure peaking is unfavorable for stability of infernal-type MHD modes, as discussed in \cite{martynov_thesis,CosteSarguet_2024}. \\
As in \cite{deVries_2018}, the ramp-down traces are shown in $\ell_{i3}-q_{95}$ space in Figure \ref{fig:qli_traces}.
The upper stability limit on $\ell_{i3}$, related to the onset of resistive MHD modes, is an increasing function of $q_{95}$ \cite{Cheng_1987}.
Evidently, a faster ramp-down leads to more significant current density peaking and hence a more elevated $\ell_{i3}(q_{95})$ trace, reducing margin with respect to MHD stability.
While the significant volume and elongation reduction foreseen for ITER and DEMO ramp-downs successfully and significantly limits the overall increase of $\ell_{i3}$, it is nevertheless important to realize that the resulting slower increase of $q_{95}$ causes a more elevated value of $\ell_{i3}$ at a given value of $q_{95}$ for the varying-shape ramp-down. 
As noted in \cite{deVries_2018}, the consequences regarding MHD stability of such prolonged dwell at relatively low $q_{95}$ values should be accounted for, assessing ideal and resistive $\beta$ limits for the internal profile dynamics consistent with the foreseen LCFS evolution.
\begin{figure}
    \centering
    \includegraphics[width=0.5\linewidth]{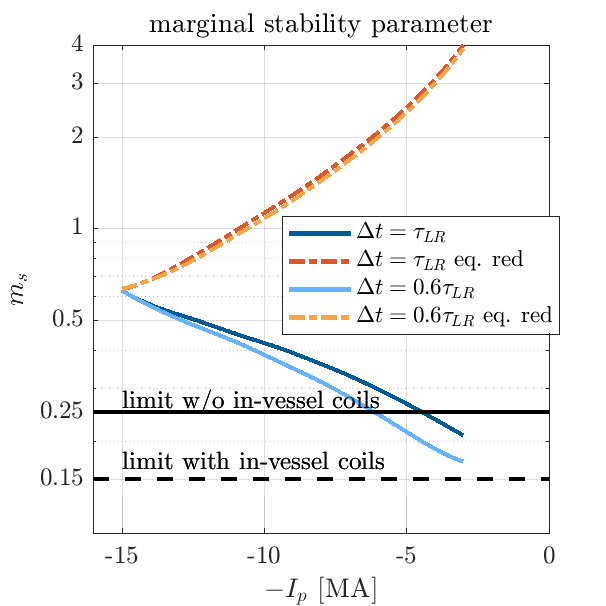}
    \caption{ITER RAPTOR ramp-down simulations over $\Delta t_{ramp-down}=\tau_{LR}$ and $\Delta t_{ramp-down}=0.6\tau_{LR}$, for constant and varying $\kappa$: $(-I_p,m_s)$ trajectories compared to the lower boundaries mentioned in \cite{deVries_2018} for control scenarios with and without in-vessel control coils.}
    \label{fig:ITER_msparam}
\end{figure}
\\
In addition to MHD stability limits, controllability over plasma shape, radial position and vertical position adds important constraints on the feasible LCFS and internal profile evolution.
As a proxy for the maximal vertical displacement that can be controlled in ITER, a marginal stability parameter has been introduced in \cite{Humphreys_2016},
\begin{equation}
m_s = \left[ \frac{1.47(1+\exp(-2\ell_{i3}+1))}{2(\kappa-1.13)}-1\right](1+0.6(\beta_p-0.1)).
\label{eq:ms}
\end{equation}
In \cite{deVries_2018}, minimum controllable $m_s$ values of $0.15$ and $0.25$ are reported, below which vertical position controllability is lost ($m_s=0.15$ relies on both in-vessel and ex-vessel coils for vertical position control, while $m_s=0.25$ considers use of ex-vessel coils only). \\
Calculating this metric for the ITER ramp-down RAPTOR simulations reported here, it is clear that a large margin to the limit values can be maintained if the elongation is reduced during ramp-down, as foreseen for ITER operation \cite{deVries_2018}, for both the slower $\Delta t_{ramp-down}=\tau_{LR}$ and the faster $\Delta t_{ramp-down}=0.6\tau_{LR}$ ramp-down simulations.
For the constant elongation ramp-down simulations, the $m_s\sim0.25$ controllability limit in presence of only ex-vessel control coils is violated in the final part of the ramp-down.
As expected, the faster ramp-down with $\Delta t_{ramp-down}=0.6\tau_{LR}$, with higher $\ell_{i3}$, reaches $m_s=0.25$ earlier.
Clearly, reduction of the elongation during ramp-down is an important actuator to ensure vertical stability, while, as we found earlier, also being an effective measure to limit the inverted current $I_{p,rev}/I_p$, at the expense of reaching a higher $\ell_{i3}$ for given $q_{95}$.
In practice, the fastest controllable ramp-down rate can be limited by the capabilities of the shape control system to deliver the desired elongation and volume reduction, as found in \cite{Gribov_2016} for ITER ramp-down scenarios.
In \cite{Gribov_2016}, a time interval of $\SI{62}{s}$ is reported for the fastest nominal plasma termination scenario from $I_p=\SI{15}{MA}$ to \SI{1.5}{MA}, maintaining a diverted configuration and the requirements regarding plasma-to-wall gap control. Interestingly, this number is very close to the $L_i/R$ time of $\tau_{LR}=\SI{63.2}{s}$, which we have proposed here as a reference time for controlled discharge termination.
\begin{figure}
    \centering
    \includegraphics[width=0.5\linewidth]{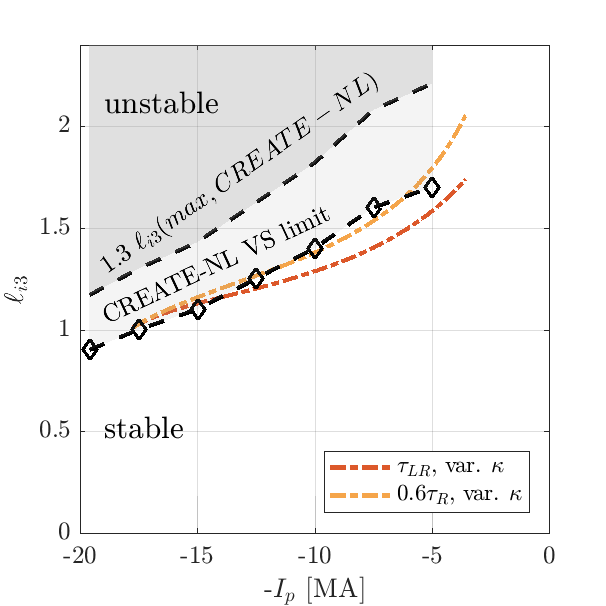}
    \caption{DEMO RAPTOR ramp-down simulations over $\Delta t_{ramp-down}=\tau_{LR}$ and $\Delta t_{ramp-down}=0.6\tau_{LR}$, for varying $\kappa$: $(-I_p,\ell_{i3})$ trajectories compared to the upper boundary mentioned in \cite{VanMulders_2024} from CREATE-NL control calculations.}
    \label{fig:DEMO_create_limit}
\end{figure}
\\
For DEMO, the slower ramp-down with $\Delta t_{ramp-down}=\tau_{LR}$ and the faster ramp-down with $\Delta t_{ramp-down}=0.6\tau_{LR}$ are compared to the CREATE-NL $\ell_{i3}(-I_p)$ stability limit reported in \cite{CREATE_NL, VanMulders_2024} in Figure \ref{fig:DEMO_create_limit}. Since this stability limit has been derived considering a reducing elongation of the LCFS during ramp-down from $\kappa=1.68$ to $1.35$, only the ramp-down simulations with varying LCFS are shown in the diagram (note however that the limit values for early ramp-down have been derived for much higher $\beta_p$ corresponding to an H-mode plasma).
For the non-linear optimization in \cite{VanMulders_2024}, $\ell_{i3}<1.3\ell_{i3}(\text{max, CREATE-NL})$ has been used as an $I_p$-dependent constraint on $\ell_{i3}$, with $\ell_{i3}(\text{max, CREATE-NL})$ the vertical position control stability limit from \cite{CREATE_NL}, arguing that some additional margin with respect to the stability limit could be assumed, with further optimization. 
The slower $\Delta t_{ramp-down}=  \tau_{LR}$ ramp-down is slightly above $\ell_{i3}(\text{max, CREATE-NL})$ at the beginning of ramp-down, but remains below this limit during most of the later ramp-down.
The faster $\Delta t_{ramp-down}=  0.6\tau_{LR}$ ramp-down evolves in closer vicinity to $\ell_{i3}(\text{max, CREATE-NL})$ and features a sharp increase of $\ell_{i3}(-I_p)$ for $I_p$ below $\sim \SI{7.5}{MA}$.
While Figure \ref{fig:DEMO_create_limit} is shown to give a rough indication of vertical stability margin, free boundary equilibrium control calculations with consistent internal profile dynamics should assess whether adequate magnetic control can be ensured for these ramp-down rates. \\
Finally it is important to note that the ramp-down simulations shown here start from Ohmic flat-top conditions.
For discharge termination from a burning plasma state, additional time should be allocated as the plasma needs to exit the burn and brought to L-mode. 
In \cite{VanMulders_2024}, RAPTOR ramp-down simulations have been performed for DEMO, starting from burning plasma conditions, including the HL transition and maintaining auxiliary heating during the L mode phase to avoid a radiative collapse in the presence of tungsten. As mentioned in Section \ref{sec:setup}, a ramp-down time of \SI{255}{s} was found to be required to avoid the formation of a significant reverse current region, which is about $50 \%$ in excess of the $L_i/R$ time of $\tau_{LR}=\SI{166.9}{s}$ found here for Ohmic plasma conditions.
\section{Analytical model to estimate $\tau_{LR}$}
\label{sec:scalinglaw}
To enable fast evaluation of the time scale $\tau_{LR}=L_i/R$ that has been proposed as the time required for controlled discharge termination (reducing $I_p$ to 20\% of $I_{p.FT}$), an analytical formula is now derived to evaluate how $\tau_{LR}$ varies across tokamaks, based on a set of physics parameters evaluated at stationary Ohmic conditions.
Let us start from the internal inductance formula $L_i=\mu_0\ell_{i3}R_0/2$ and the approximate relation for the plasma resistance $R\approx\mathcal{L}/(\langle\sigma\rangle\mathcal{A})$ with the volume-averaged conductivity $\langle\sigma\rangle$, conductor length $\mathcal{L}\approx 2 \pi R_0$ and cross-section area $\mathcal{A}\approx \pi \kappa a^2$ \cite{Sauter_2016}.
Therefore,
\begin{equation}
R\approx\frac{2R_0}{\langle\sigma\rangle \kappa a^2} 
\label{eq:R}
\end{equation}
and
\begin{equation}
\tau_{LR} = \frac{L_i}{R} \approx \mu_0 (a/2)^2 \langle\sigma\rangle \ell_{i3} \kappa.
\end{equation}
Note the similarity to the formula for the resistive time, $\tau_R=\mu_0 (a/2)^2\langle \sigma \rangle$.
Here we find $\tau_{R}/\tau_{LR}\approx 1/(\ell_{i3}\kappa)$, which evaluated for the four tokamaks, approximates within about 10\% the ratios $\tau_{R}/\tau_{LR}$ obtained in Table \ref{tab:flattop} for TCV, JET, ITER and DEMO. \\
To approximate the volume-averaged conductivity $\langle\sigma\rangle$, we assume $\langle\sigma\rangle \ \approx \overline{ \sigma_{Sptz}}\cdot  \overline{ \sigma_{neo}/\sigma_{Sptz}}$, with average Spitzer conductivity $\overline{\sigma_{Sptz}}$ and an average neoclassical correction factor $\overline{ \sigma_{neo}/\sigma_{Sptz}}$.
We follow the formulas reported in \cite{Sauter_1999} to define the average Spitzer conductivity $\overline{ \sigma_{Sptz}}$ as
\begin{equation}
\overline{ \sigma_{Sptz}}=1.9014\ 10^{4} \frac{\langle T_e  \rangle_{vol}^{3/2}}{Z_{eff}N(Z_{eff})\overline{\ln\Lambda_e}},
\end{equation}
with 
\begin{equation}
N(Z_{eff})=0.58+\frac{0.74}{0.76+Z_{eff}}
\end{equation}
and
with
\begin{equation}
\overline{\ln\Lambda_e} =31.3-\ln \left( \frac{\sqrt{\langle n_e  \rangle_{vol}}}{\langle T_e  \rangle_{vol}} \right).
\end{equation}
We then derive a formula for the average neoclassical correction factor $\overline{ \sigma_{neo}/\sigma_{Sptz}}$, by introducing $\overline{ \epsilon }=\epsilon/2$, $\overline{ \kappa }=(1+\kappa)/2$, $\overline{ \delta }=\delta/2$ and $\overline{ q}=(1+q_{95})/2$, which are essentially mean values of the shaping variables and the safety factor at the magnetic axis and near the LCFS. 
Using the formulas found in \cite{Sauter_1999}, we then obtain
\begin{equation}
\overline{ \sigma_{neo}/\sigma_{Sptz}} = 1-\left(1+\frac{0.36}{Z_{eff}}\right)\overline{ f_t^{33} }+\frac{0.56}{Z_{eff}}\overline{ f_t^{33} }^2 - \frac{0.23}{Z_{eff}} \overline{ f_t^{33} }^3   , 
\end{equation}
with
\begin{equation}
\overline{ f_t^{33} }= \frac{\overline{ f_{t} }}{1+(0.55-0.1\overline{f_t})\sqrt{\overline{ \nu_{e*} }}+0.45(1-\overline{f_t})\overline{ \nu_{e*} }/Z_{eff}^{3/2}}.
\label{eq:ft33}
\end{equation}
In eq. (\ref{eq:ft33}), we use an average collisionality $\overline{ \nu_{e*}}$, 
\begin{equation}
\overline{ \nu_{e*}}=6.921\ 10^{-18} \frac{\overline{q} R_0 \langle n_e  \rangle_{vol}Z_{eff}\overline{\ln\Lambda_e}}{\langle T_e \rangle_{vol}^2 \overline{ \epsilon }^{3/2}},
\end{equation}
and an average trapped particle fraction, which we define, based on the formula in \cite{Sauter_2016}, as
\begin{equation}
\overline{ f_t } = 1- \frac{1-\overline{\epsilon_{eff}}}{1+2\sqrt{\overline{\epsilon_{eff}}}}\sqrt{\frac{1-\overline{ \epsilon }}{1+\overline{ \epsilon }}},
\end{equation}
with
\begin{equation}
\overline{\epsilon_{eff}} = 0.67(1-1.4\overline{\delta}|\overline{\delta}|)\overline{\epsilon}.
\end{equation}
Using the $\tau_{LR}$ value we found for ITER as a reference $\tau_{LR}|_{ITER}$, we can then write a formula for how $\tau_{LR}$ scales as a function of $a$, $\langle T_e \rangle_{vol}$, $\langle \sigma_{neo}/\sigma_{Sptz}\rangle$, $\ell_{i3}$ and $\kappa$,
\begin{equation}
    \tau_{LR} = \tau_{LR}|_{ITER}  \left( \frac{a}{a|_{ITER}} \right)^2   \left( \frac{\langle T_e \rangle_{vol}}{\langle T_e \rangle_{vol}|_{ITER}} \right)^{3/2} \frac{\overline{ \sigma_{neo}/\sigma_{Sptz}}}{\overline{ \sigma_{neo}/\sigma_{Sptz}}|_{ITER}}\frac{\ell_{i3}}{\ell_{i3}|_{ITER}} \frac{\kappa}{\kappa|_{ITER}}
    \label{eq:scalinglaw}
\end{equation}
with $\tau_{LR}|_{ITER}=\SI{63.2}{s}$, $a|_{ITER}=\SI{2}{m}$, $\langle T_e \rangle_{vol}|_{ITER}=\SI{1.47}{keV}$, $\overline{ \sigma_{neo}/\sigma_{Sptz}}|_{ITER}=0.66$, $\ell_{i3}|_{ITER}=0.87$,  and $\kappa|_{ITER}=1.80$, assuming $Z_{eff}=1.5$.
Since cross-machine variation of $\langle \sigma_{neo}/\sigma_{Sptz}\rangle$, $\ell_{i3}$ and $\kappa$ is generally speaking modest, the dominant dependence in this scaling law is $\tau_{LR} \sim a^2 \langle T_e \rangle_{vol}^{3/2}$.
Evaluating eq. (\ref{eq:scalinglaw}) for the three other tokamaks previously considered, with $\langle T_e \rangle_{vol}$ based on the flat-top segment of the RAPTOR simulations, using the values from Table \ref{tab:flattop}, we obtain $\tau_{LR}=\SI{0.030}{s}$ for TCV, $\tau_{LR}=\SI{2.53}{s}$ for JET, $\tau_{LR}=\SI{167.7}{s}$ for DEMO, all within 12\% of the $\tau_{LR}$ values found from the RAPTOR simulations reported in Table \ref{tab:flattop}.  
Eq. (\ref{eq:scalinglaw}) allows to estimate $\tau_{LR}$, and hence the (minimum) timescale proposed in this paper for controlled discharge termination in an Ohmic plasma, for a given $\langle T_e \rangle_{vol}$.
Note however that eq. (\ref{eq:scalinglaw}) can also be used inversely: for a given desired ramp-down time, the formula provides an estimate of the maximum allowable $\langle T_e \rangle_{vol}$ under Ohmic flat-top conditions, implying that the desired ramp-down time cannot be achieved in a controlled way if the volume-averaged electron temperature is higher than the value obtained with eq. (\ref{eq:scalinglaw}).
\begin{figure}
    \centering
    \includegraphics[width=1.\linewidth]{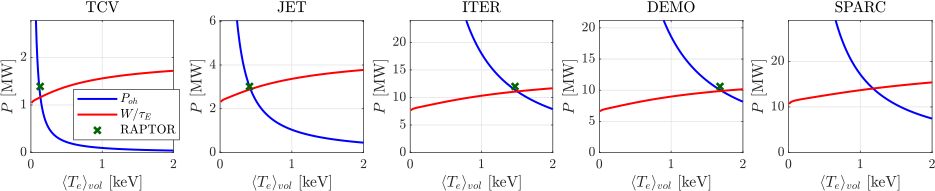}
    \caption{Ohmic power $RI^2$ and radial transport power sink $W/\tau_E$ as a function of $\langle T_e\rangle_{vol}$, evaluated for $Z_{eff}=1.5$, $f_{Gw}=1$ and $H_{98y,2}=0.4$ for TCV, JET, ITER and SPARC. The crosses indicate the $(\langle T_e \rangle_{vol},P_{oh})$ pairs based on the flat-top segment of the RAPTOR simulations described in Section \ref{sec:model_setup}.}
    \label{fig:powerbalance}
\end{figure}
\\
In the RAPTOR simulations described in Section \ref{sec:model_setup}, we have assumed $Z_{eff}=1.5$ for the effective charge, $f_{Gw}=\langle n_e \rangle_{vol}/(\pi a^2)=1$ for the Greenwald density fraction and $H_{98y,2}=\tau_E/\tau_{E,98y,2}=0.4$ for the confinement factor with respect to the IPB98(y,2) scaling law.
To explore how $\langle T_e \rangle_{vol}$ varies for different tokamaks, or for different assumptions regarding $Z_{eff}$, $f_{Gw}$ and $H_{98y,2}$, we can write a simple zero-dimensional power balance equation for the evolution of the thermal energy $W_{th}$,
\begin{equation}
    \frac{dW_{th}}{dt}=P_{oh}-\frac{W_{th}}{\tau_E},
\end{equation}
with Ohmic power $P_{oh}=RI^2$ and the power sink due to radial transport $-W_{th}/\tau_E$, with the energy confinement time $\tau_E$, neglecting energy losses through radiation.
The equilibrium operating point satisfies $dW_{th}/dt=0$, hence
\begin{equation}
    RI^2=\frac{W_{th}}{\tau_E}.
\end{equation}
Assuming ions and electrons at equal temperature and density, the thermal energy can be written approximately as $W_{th}\approx3e\langle T_e\rangle_{vol}\langle n_e  \rangle_{vol}V$, with $e=1.60218\ 10^{-19}$ for the conversion from eV to J and V the volume enclosed by the LCFS. 
The energy confinement time can be written as $\tau_E= H_{98y,2} \tau_{E,98y,2}$, with $\tau_{E,98y,2}=0.0562I_p^{0.93}B_0^{0.15}P_l^{-0.69}\langle n_e \rangle_{vol}^{0.41}M^{0.19}R_0^{1.97}\epsilon^{0.58}\kappa^{0.78}$ \cite{ITER_TC_1999}, with $M$ the average isotope mass number and $P_l$ the loss power crossing the LCFS.
Given our assumptions, we can equate the loss power to the Ohmic power, i.e. $P_l=RI^2$.
Both left-hand-side $RI^2$ and right-hand-side $W_{th}/\tau_E$ of the power balance equation depend on $\langle T_e\rangle_{vol}$.
In Figure \ref{fig:powerbalance}, we show $RI^2$ and $W_{th}/\tau_E$, calculating $R$ based on eq. (\ref{eq:R}) (with the formulas for average Spitzer conductivity $\overline{\sigma_{Sptz}}$ and average neoclassical correction factor $\overline{ \sigma_{neo}/\sigma_{Sptz}}$ defined above) and maintaining the assumptions $Z_{eff}=1.5$, $f_{Gw}=1$, $H_{98y,2}=0.4$.
The resulting curves are shown for TCV, JET, ITER and DEMO, as well as for the SPARC full-field scenario reported in \cite{Creely_2020} ($B_0=\SI{12.2}{T}$, $R_0=\SI{1.85}{m}$, $a=\SI{0.57}{m}$, $I_{p,FT}=\SI{8.7}{MA}$, $\epsilon=0.31$, $\kappa=1.97$, $\delta=0.54$, $q_{95}=3.05$, $V\approx2\pi R_0 \pi a^2\kappa=\SI{23.4}{m^3}$\footnote{For the SPARC $\tau_{LR}$ estimates, we assume $\ell_{i3}=\ell_{i3}|_{ITER}=0.87$.}).
The intersection point of both curves in Figure \ref{fig:powerbalance} indicates the stationary operating points for the various tokamaks at the flat-top $I_{p,FT}$ (these operating points are stable since the Ohmic power decreases while the power sink due to radial transport increases for a perturbation where $\langle T_e\rangle_{vol}$ increases).
For TCV, JET, ITER and DEMO, the $(\langle T_e \rangle_{vol},P_{oh})$ pairs based on the flat-top segment of the RAPTOR simulation are superimposed in Figure \ref{fig:powerbalance}. The close proximity of these points to the stationary operating points corroborates the robustness of the zero-dimensional power balance method proposed here.
\begin{table}[htbp]
\centering
\caption{$\langle T_e\rangle_{vol}$ and $\tau_{LR}$ calculated based on the proposed analytical model, for TCV, JET, ITER, DEMO, SPARC, for varying assumptions regarding $Z_{eff}$, $f_{Gw}$ and $H_{98y,2}$.}
\label{tab:sensitivity}
\begin{tabular}{l|cc|cc|cc|cc|cc}
\toprule
 & \multicolumn{2}{c|}{TCV} 
 & \multicolumn{2}{c|}{JET} 
 & \multicolumn{2}{c|}{ITER} 
 & \multicolumn{2}{c|}{DEMO} 
 & \multicolumn{2}{c}{SPARC} \\
\midrule
$Z_{eff}; f_{Gw}; H_{98y,2}$
 & $\langle T_e\rangle_{vol}$ & $\tau_{LR}$ 
 & $\langle T_e\rangle_{vol}$ & $\tau_{LR}$ 
 & $\langle T_e\rangle_{vol}$ & $\tau_{LR}$ 
 & $\langle T_e\rangle_{vol}$ & $\tau_{LR}$ 
 & $\langle T_e\rangle_{vol}$ & $\tau_{LR}$ \\
 & [keV] & [s]
 & [keV] & [s] 
 & [keV] & [s] 
 & [keV] & [s] 
 & [keV] & [s] \\
\midrule
1.5; 1.0; 0.4 & 0.133 & 0.031 & 0.425 & 2.62 & 1.501 & 64.7 & 1.704 & 171 & 1.178 & 4.69 \\
3.0; 1.0; 0.4 & 0.150 & 0.037 & 0.475 & 3.19 & 1.666 & 80.4 & 1.888 & 213 & 1.316 & 5.74 \\
1.5; 0.5; 0.4 & 0.178 & 0.044 & 0.572 & 3.53 & 2.024 & 87.9 & 2.295 & 233 & 1.587 & 6.35 \\
1.5; 1.0; 0.8 & 0.218 & 0.061 & 0.704 & 4.98 & 2.489 & 123 & 2.820 & 326 & 1.953 & 8.94 \\
3.0; 0.5; 0.8 & 0.328 & 0.104 & 1.053 & 8.41 & 3.698 & 214 & 4.185 & 570 & 2.919 & 15.2 \\
\hline
\end{tabular}
\end{table}
\\
Let us now use the analytical formulas presented in this section to estimate $\tau_{LR}$ for a range of different assumptions regarding $Z_{eff}$, $f_{Gw}$ and $H_{98y,2}$ for TCV, JET, ITER, DEMO and SPARC (note that the change of $\ell_{i3}$ in eq. (\ref{eq:scalinglaw}) due to the variation of these variables is not accounted for).
Compared to the default assumptions chosen in Section \ref{sec:model_setup}, i.e. $Z_{eff}=1.5$, $f_{Gw}=1$, $H_{98y,2}=0.4$, Table \ref{tab:sensitivity} includes sensitivity studies with respectively increased $Z_{eff}=3.0$, reduced $f_{Gw}=0.5$ and increased $H_{98y,2}=0.8$, as well as a combination of all of these in the final row of the table.
The values of $\tau_{LR}$ found for the default assumptions, in the first row of Table \ref{tab:sensitivity}, are all within 10\% of the values obtained with the corresponding RAPTOR simulation in Section \ref{sec:model_setup}.
Comparing the various rows in Table \ref{sec:model_setup}, we find that a doubling of effective charge to $Z_{eff}=3$ leads to an increase of $\tau_{LR}$ by about 20\%-25\%; halving the density to $f_{Gw}=0.5$ leads to an increase of $\tau_{LR}$ by about 35\%-40\%; and doubling the confinement factor to $H_{98y,2}=0.8$ leads to an increase of $\tau_{LR}$ by about 90\%-97\%. Combining all three effects leads to an increase of $\tau_{LR}$ by about a factor 3.
Note that the controlled ramp-down time for the SPARC tokamak, with $\tau_{LR}$ in the range \SI{4.69}{s}-\SI{15.2}{s}, is in good agreement with the \SI{12}{s} ramp-down reported in \cite{Creely_2020}.
\section{Conclusion}
Transport simulations for Ohmic plasmas, performed for tokamaks spanning a wide range of major radius and toroidal magnetic field, indicate that a plasma current ramp-down over a time interval $\tau_{LR}=L_i/R=I_p^2L_i/P_{oh}$, evaluated for nominal flat-top conditions, avoids significant reversal of the plasma current in the edge region of the plasma. 
The simulations have been performed for a linearly reducing $I_p$ from the flat-top value $I_{p,FT}$ down to $0.2I_{p,FT}$.
We propose $\Delta t=\tau_{LR}$ as a reference for the minimum time required for controlled discharge termination, yielding \SI{0.033}{s} for TCV, \SI{2.87}{s} for JET, \SI{63.2}{s} for ITER and \SI{166.9}{s} for DEMO. 
\\
A reduced ramp-down time equal to $0.6\tau_{LR}$ (comparable to the resistive time $\tau_R=\mu_0 (a/2)^2\langle \sigma(t) \rangle$, evaluated for nominal flat-top conditions) is found to cause significant reversal of the plasma current density in the outer plasma core, carrying about $20\%$ to $40\%$ of the plasma current in the direction opposite to $I_p$.
While a slower ramp-down with $\Delta t=\tau_{LR}$ leads to a final internal inductance $\ell_{i3}$ around 2, the faster ramp-downs with $\Delta t=0.6\tau_{LR}$ cause significant further current density peaking, reaching $\ell_{i3}\sim3$, a value beyond what is typically observed in present-day controlled ramp-down experiments \cite{deVries_2018}.
This current density reversal is successfully avoided through the significant reduction of the plasma cross-section and elongation during ramp-down, as foreseen for ITER and DEMO ramp-downs, also maintaining $\ell_{i3}<2$. 
However the feasibility of such rapid shape control should be assessed, as in \cite{Gribov_2016}. 
Furthermore, implications regarding MHD stability should be studied, since ramp-downs with significant cross-section contraction follow a higher path in $\ell_{i3}-q_{95}$ space, in closer vicinity to the upper stability limit \cite{Cheng_1987}.
While the obtained ramp-down scenarios have been tested against some simple vertical stability metrics available in the literature \cite{Humphreys_2016, Humphreys_2016}, magnetic control calculations, accounting for vertical stability control as well as shape control, considering the obtained internal profile dynamics, remain invaluable for proper assessment of controlled termination scenarios.
Furthermore, additional time should be accounted for to exit the plasma burn and bring the plasma to L-mode, even if these transitions can be performed during the $I_p$ ramp-down.
As auxiliary heating is likely required to avoid radiative collapse in presence of intrinsic tungsten impurities from the first wall, increasing plasma conductivity with respect to the Ohmic conditions assumed here, a further prolongation of the effective minimum controllable ramp-down time can be expected. 
\\
Note that the proposed reference time for controlled discharge termination $\tau_{LR}=L_i/R$ can easily be evaluated in real-time during the flat-top phase, since for constant $I_p$, $P_{oh}=U_{pl,b}I_p$ (see eq. (\ref{eq:Poynting})), and hence $\tau_{LR}=I_pL_i/U_{pl,b}$ with $L_i=\mu_0R_o \ell_{i3}/2$, where $I_p$, $U_{pl,b}$ and $\ell_{i3}$ can be obtained from magnetic equilibrium reconstruction.
Finally, the metric could also be used in systems codes as a first-order estimate for realistic controlled ramp-down times.
A simple analytical model has been proposed to evaluate $\tau_{LR}$ based on engineering parameters, allowing to assess how the controlled ramp-down time scales across different machines and across scenarios with varying effective charge, Greenwald density and confinement quality.
\\
Future research should assess the stability of the fast ramp-down regime with significant reversal of the outer current density profile, both theoretically and experimentally, to improve understanding of the implications of negative edge current density, peaked current density and pressure profiles with respect to vertical position control, shape control and MHD stability. 

\appendix
\section{Mathematical derivation: $L_i$ and $\psi_{LCFS}-\psi_{axis}$}
\label{sec:Li_deriv}
Let us start by writing the volume enclosed by a flux surface as the integral
\begin{equation}
V = \int dV = \int Rd\phi \frac{d\psi}{|\nabla\psi|}d\ell_p=\int d\psi\oint\frac{d\ell_p}{B_p},
\label{eq:volint}
\end{equation}
with $d\ell_p$ an infinitesimal length in the poloidal plane along a flux surface and $B_p=|\nabla\psi|/(2\pi R)$ the magnitude of the local poloidal field.
The total energy of the magnetic field due to $I_p$ inside the LCFS, previously written as $L_{i}I_p^2/2$, can be written in terms of the volume integral of $B_p^2/(2\mu_0)$, i.e.
\begin{equation}
\frac{L_{i}I_p^2}{2} = \frac{1}{2\mu_0}\int B_p^2 dV =\frac{1}{2\mu_0}\int d\psi\oint B_pd\ell_p,
\end{equation}
and applying Amp\`{e}re's law
\begin{equation}
L_{i}I_p^2 =\int d\psi I_{p,int}(\psi),
\end{equation}
with $I_{p,int}(\psi)$ the toroidal current inside a flux surface.
Therefore,
\begin{equation}
L_{i}I_p =\int d\psi I_{p,int}(\psi)/I_p,
\label{eq:Iphi_int}
\end{equation}
Since the integrand $I_\phi(\psi)/I_p$ evolves from 0 at the magnetic axis to 1 at the LCFS, we obtain,
\begin{equation}
L_{i}I_p \sim \psi_{LCFS}-\psi_{axis},
\end{equation}
justifying the normalization of $\psi_{LCFS}-\psi_{mag.\ axis}$ used in Figure \ref{fig:diffpsi}.
Note furthermore that the integral eq. (\ref{eq:Iphi_int}) illustrates that for a given plasma current $I_p$, the internal inductance increases when the current density is more peaked.
\ack
The authors would like to acknowledge the insightful discussions with Sergei Medvedev and Peter de Vries. 
\bibliographystyle{unsrt}
\bibliography{References}

\end{document}